\documentclass[prb,showpacs,amsmath,amssymb,superscriptaddress,floatfix,twocolumn]{revtex4}

\usepackage[english]{babel}
\usepackage{graphicx}
\usepackage{times}

\begin{document}

\title{Anomaly in the heat capacity of Kondo superconductors}

\author{Rok \v{Z}itko}
\affiliation{Institute for Theoretical Physics, University of G\"ottingen,
Friedrich-Hund-Platz 1, D-37077 G\"ottingen, Germany}
\affiliation{J. Stefan Institute, Jamova 39, SI-1000 Ljubljana, Slovenia}

\author{Thomas Pruschke}
\affiliation{Institute for Theoretical Physics, University of G\"ottingen,
Friedrich-Hund-Platz 1, D-37077 G\"ottingen, Germany}

\date{\today}

\begin{abstract}
Using numerical renormalization group we study thermodynamic properties of a
magnetic impurity described by the Anderson impurity model in a
superconducting host material described by the BCS Hamiltonian. When the
Kondo temperature in the normal state, $T_K$, is comparable to the critical
temperature of the superconducting transition, $T_c$, the magnetic doublet
state may become degenerate with the Kondo singlet state, leading to a $\ln
3$ peak in the temperature dependence of the impurity contribution to the
entropy. This entropy increase translates into an anomalous feature in the
heat capacity which might have already been experimentally observed.
\end{abstract}


\maketitle

\newcommand{\vc}[1]{\boldsymbol{#1}}
\newcommand{\ket}[1]{|#1\rangle}
\newcommand{\braket}[1]{\langle #1 \rangle}
\newcommand{\korr}[1]{\langle\langle #1 \rangle\rangle_\omega}
\newcommand{\cG}{\mathcal{G}}
\newcommand{\erfc}[1]{\mathrm{erfc}#1}
\newcommand{\sgn}[1]{\mathrm{sgn}#1}

\renewcommand{\Im}{\mathrm{Im}}
\renewcommand{\Re}{\mathrm{Re}}

\newcommand{\figw}{7cm}

\section{Introduction}

After many decades of research, conventional superconductors are still being
actively studied both theoretically and experimentally \cite{aynajian2008}.
An important experimental technique to study long-range-ordered states
consists in determining what destroys that order \cite{balatsky2006}.
Superconductivity can be suppressed, for example, by doping the host
material with magnetic impurities. Each magnetic impurity locally perturbs
the superconducting order; this can lead to a significant cumulative effect
and ultimately, at large concentrations, to a complete suppression of the
superconductivity \cite{balatsky2006}.

In the normal state of a metal, the impurity magnetic moment tends to be
screened by the Kondo effect \cite{hewson}. In the superconducting state,
however, the gap in the spectrum implies that there are no low-energy
electrons to perform the screening \cite{balatsky2006}. The behavior
thus depends on the ratio between the Kondo temperature $T_K$ and the
critical temperature $T_c$ for superconducting transition. In the $T_K \ll
T_c$ limit, there is no Kondo effect since the gap is fully formed on the
relevant energy scale of $T_K$, while in the $T_K \gg T_c$ limit, Kondo
physics does not affect the superconducting transition since the impurities
are already fully nonmagnetic (Kondo screened) on the scale of $T_c$. The
non-trivial regime is thus $T_K \sim T_c$, when both effects compete. To
study this competition and demonstrate the existence of the Kondo effect in
superconductors, one can study, for example, the effects of the impurities
on the density of states \cite{zmha, dumoulin1975} or the thermodynamic
\cite{armbruster1974, takeuchi1978} and transport properties
\cite{dreyer1982}. With advances in the tunneling spectroscopy, focus has
recently shifted to dynamic properties, i.e. variation of the local density
of states induced by the presence of the impurity \cite{yazdani1997,
fischer2007, balatsky2006}.

The effect of impurities on the host is the central focus in the field of
quantum impurity physics \cite{hewson, bulla2008}. Studying thermodynamics
of an impurity model is a common practice in this field, as one can extract
many essential aspects of the impurity behaviour (its magnetic properties,
effective degrees of freedom, etc.).  A reliable theoretical tool is the
numerical renormalisation group (NRG) \cite{wilson1975, krishna1980a,
bulla2008}. The NRG has been applied, for example, to study the sub-gap
bound states in isotropic \cite{satori1992} and anisotropic
\cite{yoshioka1998} Kondo model and in the Anderson model
\cite{yoshioka2000} in both conventional and unconventional
\cite{koga2002sc, matsumoto2001} superconductors. Further NRG calculations
focused on dynamical properties such as the spectral function
\cite{sakai1993, vojta2001sc, vojta2002stm, bauer2007} and more recently on
the transport properties of quantum dots attached to superconducting leads
\cite{oguri2004josephson, choi2004, karrasch2008, tanaka2007normal,
tanaka2007josephson}. Only few results of NRG calculations for thermodynamic
quantities, such as the impurity contribution to the magnetic susceptibility
and the entropy, have been published so far \cite{vojta2001sc,
matsumoto2002}. In this paper we present a detailed investigation of these
thermodynamic properties of an Anderson impurity in a BCS superconducting
host. We find features in the thermodynamic behavior of magnetically doped
superconductors (also known as the ``Kondo superconductors'') in the dilute
doping limit which appear to have been overlooked in theoretical treatments,
but may have been already experimentally observed.

\section{Model and method}

A conceptually proper way of studying the thermodynamics of a magnetic
impurity in a superconducting host would start from a microscopic
Hamiltonian for a conduction bath with explicit attractive electron-electron
interaction (quartic) terms. The NRG can only be applied to non-interacting
continuum Hamiltonians or, more accurately, to Hamiltonians with at most
quadratic (mean-field) terms. We thus focus on a mean-field BCS Hamiltonian,
keeping in mind that this is only an effective low-temperature Hamiltonian
and that its use in the NRG to compute temperature-dependent thermodynamic
quantities is associated with a certain ambiguity. In a superconductor, the
superconducting order is namely established gradually as the temperature is
reduced below $T_c$. In this work, we will make the following approximation:
\begin{equation}
\braket{O}(T)=\braket{O(\Delta=\Delta(T))}(T),
\end{equation}
i.e. at temperature $T$ we take results computed in a calculation with a
constant $\Delta$ chosen such that $\Delta=\Delta(T)$. We will use a
phenomenological approximation for the temperature dependence for the
superconductor gap that is correct near $T=0$ and $T=T_c$:
\begin{equation}
\Delta(T) \approx \delta_{sc} T_c \tanh\left[ \frac{\pi}{\delta_{sc}}
\sqrt{a \frac{\delta C}{C_N} \left( \frac{T_c}{T}-1 \right) }
\right],
\end{equation}
with $\delta_{sc}=1.76$, $a=2/3$, $\delta C/C_N=1.43$.

Even for a single impurity, the order parameter becomes a spatially varying
function $\Delta(\vc{r})$ \cite{flatte1997prl, balatsky2006}. In this work,
we disregard any spatial variation of the order parameter. In fact, we
neglect any variation of $\Delta$, considering it as a fixed (but
doping-concentration dependent) quantity. In ppm to percent doping range,
the distance between the impurities is much smaller than the coherence
length in conventional superconductors, thus the assumption of spatially
constant order parameter is justified when impurity-position-averaged
thermodynamic properties are considered \cite{jarrell1990}.

The impurity contribution to an expectation value is commonly defined by
taking the difference between the values for a doped and for a clean system:
\begin{equation}
\braket{O}_\mathrm{imp} = \braket{O} - \braket{O}_0
\end{equation}
The impurity contribution can be further decomposed into two parts: the
``intrinsic'' local effect of an impurity and the average effect of all the
impurities on the bulk pairing order parameter $\Delta$ (which needs to be
calculated self-consistently). It is the first contribution that we are
interested in in this work.

We thus consider a single Anderson impurity embedded in a superconducting
host described by the Hamiltonian $H=H_\mathrm{bath} + H_\mathrm{imp} +
H_\mathrm{hyb}$. We model the fermionic bath with a BCS model:
\begin{equation}
H_\mathrm{bath} = \sum_{k,\sigma} \epsilon_k c^\dag_{k\sigma} c_{k\sigma}
-\Delta \sum_k \left( c^\dag_{k\uparrow} c^\dag_{-k\downarrow}
+ c_{-k\downarrow} c_{k\uparrow} \right),
\end{equation}
where $c_{k\sigma}$ are conduction band operators, $\epsilon_k$
single-particle energies and $\Delta$ the BCS order parameter. For
simplicity we assume that the band has a flat density of states in the normal
state, $\rho(\omega)=1/2D$, where $2D$ is the width of the conduction band.
In the original BCS approach, the order parameter $\Delta$ has to be
determined self-consistently through $\Delta = V \sum_k \braket{
c_{-k\downarrow} c_{k\uparrow}}$, where $V$ is some phenomenological
effective coupling constant and the summation is to be cut off by the Debye
energy $\omega_D$. In numerical calculations, however, we consider $\Delta$
to be some given parameter. The magnetic impurity Hamiltonian has the
usual form
\begin{equation}
H_\mathrm{imp} = \left( \epsilon+\frac{U}{2} \right) (n-1)
+ \frac{U}{2} (n-1)^2,
\end{equation}
where $\epsilon$ is the impurity energy level, $U$ the electron-electron
repulsion and $n$ the level occupancy operator, $n=\sum_\sigma d^\dag_\sigma
d_\sigma$. Finally, the hybridisation Hamiltonian is
\begin{equation}
H_\mathrm{hyb} = V \sum_{k\sigma} \left( d^\dag_\sigma c_{k\sigma} +
\text{H.c.} \right).
\end{equation}
The coupling to the conduction band is characterized by the hybridisation
strength $\Gamma=\pi \rho V^2$.

Due to strong exchange scattering of conduction band electrons on magnetic
impurities \cite{kondo1964, hewson}, magnetic impurity problems need to be
solved using non-perturbative methods such as NRG. We performed the
calculations with a discretization parameter $\Lambda=3$, four values of
the twist parameter $z$ and the truncation energy cutoff of
$E_\mathrm{cutoff}=10 \omega_N$, where $\omega_N$ is the characteristic
energy-scale at the $N$-th NRG iteration. Only spin $\mathrm{SU}(2)$
symmetry can be explicitly used to simplify calculations with models of this
class. 

\section{Numerical results}

\begin{figure}[htbp]
\includegraphics[width=8cm,clip]{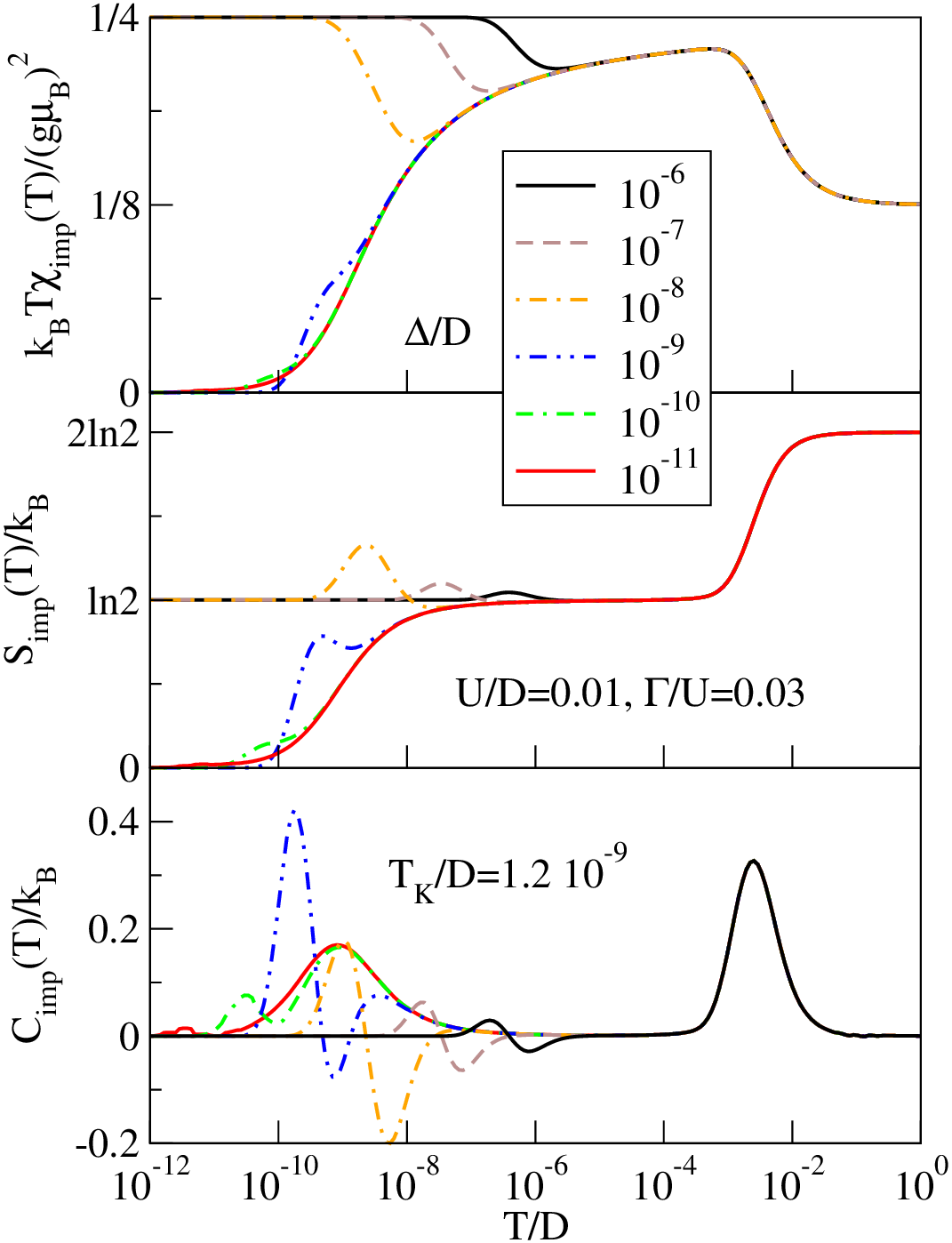}
\caption{Thermodynamic properties of an Anderson impurity
in a BCS superconductor, assuming a fixed value of $\Delta$.
The system is particle-hole symmetric: $\epsilon=-U/2$.
}
\label{fig_a}
\end{figure}

We present results for the following thermodynamic quantities: 1) the
temperature-dependent impurity contribution to the magnetic susceptibility,
$\chi_\mathrm{imp}(T) = (g\mu_B)^2/(k_B T) \left( \braket{S_z^2} -
\braket{S_z^2}_0 \right)$, where the subscript $0$ refers to the clean
system (where the Hamiltonian is simply the band Hamiltonian
$H_\mathrm{band}$ without the impurity terms), $S_z$ is the total system
spin, $g$ the electronic gyromagnetic factor, $\mu_B$ the Bohr magneton and
$k_B$ the Boltzmann's constant;

2) the temperature-dependent impurity contribution to the entropy,
$S_\mathrm{imp}(T) = (E-F)/T - (E-F)_0/T$, where $E = \braket{H}$ and
$F=-k_BT \ln \mathrm{Tr} \left( e^{-H/k_B T} \right)$;

3) the temperature-dependent impurity contribution to the heat capacity,
$C_\mathrm{imp}(T) = k_B \left( \braket{H^2} - \braket{H}^2 \right) - k_B
\left( \braket{H^2}-\braket{H}^2 \right)_0$.

\begin{figure}
\includegraphics[width=8cm,clip]{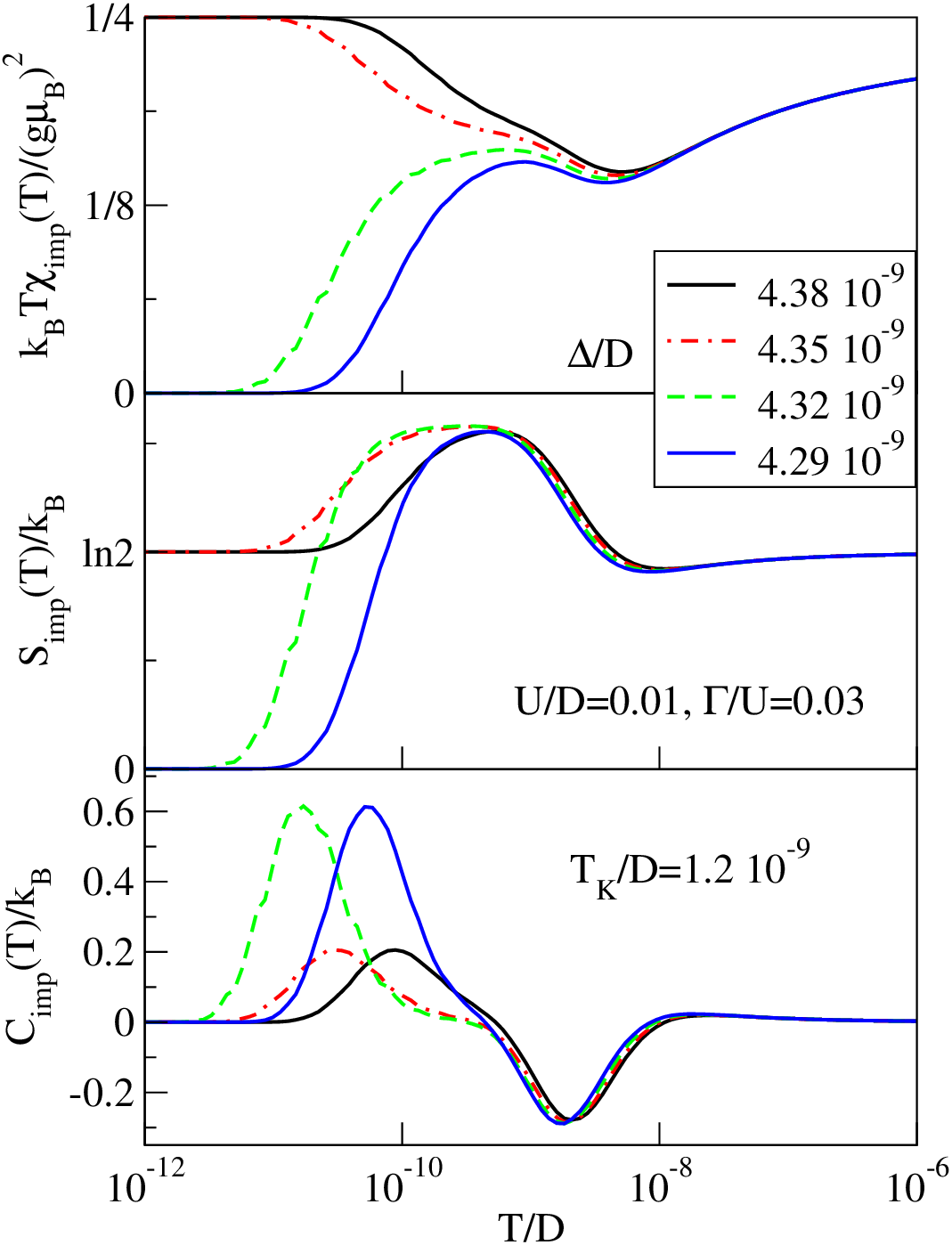}
\caption{Close-up on the low-temperature region in the transition regime.}
\label{fig_b}
\end{figure}

We first study the case of a constant gap parameter $\Delta$, i.e. a
fictitious gap parameter which does not change with the temperature. We
computed thermodynamic properties for a range of $\Delta$ at fixed Kondo
temperature, Fig.~\ref{fig_a}. (We use Wilson's definition of $T_K$, i.e.
$k_B T_K \chi(T_K)/(g\mu_B)^2=0.07$ in the normal state.) For $\Delta \ll
T_K$, we obtain the well-known results for an Anderson impurity in a normal
metal exhibiting the formation of local moment at $T \sim U$ and the Kondo
screening at $T \sim T_K$, with the system ultimately ending up in the
strong-coupling fixed point \cite{krishna1980a}. The presence of the gap is
hardly reflected in the impurity thermodynamic properties, although we
observe a small feature at $T \sim \Delta$, when Kondo screening is
interrupted and the remaining residual entropy is released. For $\Delta \gg
T_K$, the Kondo screening is again interrupted at $T \sim \Delta$, but this
time the system ends up in the local-moment fixed point rather than the
strong-coupling fixed point: the impurity spin decouples entirely, so we
obtain Curie-Weiss behaviour at low temperatures and a $\ln 2$ residual
entropy. We bring attention to the small bump in the impurity entropy above
$\ln 2$ at $T \sim \Delta$, which is mirrored in an S-shaped feature in the
impurity heat capacity. For $\Delta \sim T_K$, the behavior is similar to
that in the $\Delta \gg T_K$ regime, but the entropy increase at $T \sim
\Delta$ becomes increasingly strong as the transition point is approached.
In Fig.~\ref{fig_b} we show a close-up on the transition region. Near the
transition, the entropy rises to exactly $\ln 3$, forming a relatively broad
plateau: at this point the magnetic doublet and the Kondo singlet states are
degenerate, although this fixed point is unstable. The transition is clearly
first-order, since the quantum numbers of the ground state change. We
present results only for the particle-hole symmetric case, as asymmetry does
not affect the results qualitatively.

\begin{figure}[htbp]
\includegraphics[width=8cm,clip]{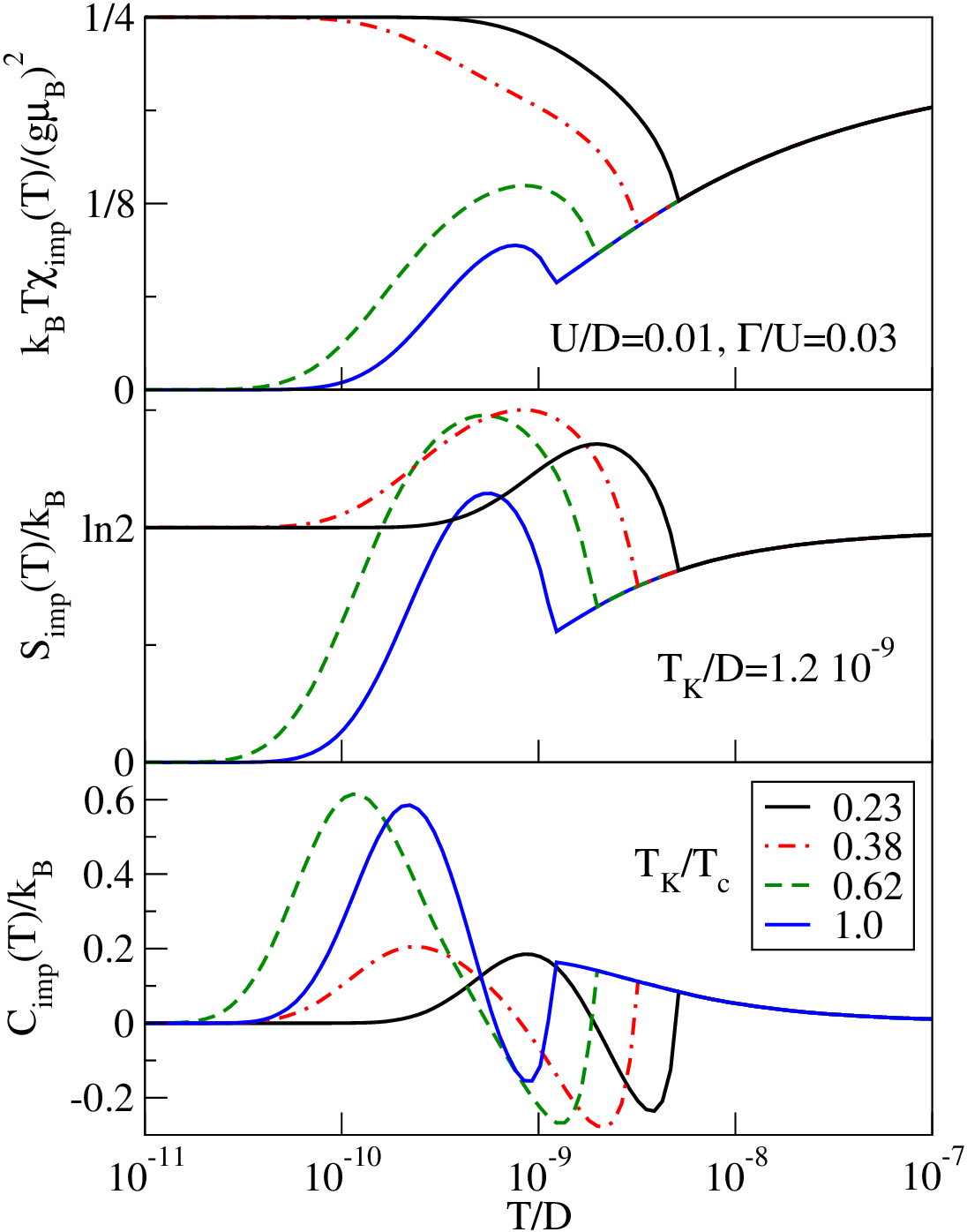}
\caption{Thermodynamic properties of an Anderson impurity
in a BCS superconductor, taking into account the temperature
dependence of the gap.}
\label{fig_c}
\end{figure}

We now turn to the case of BCS superconductors, where the gap is temperature
dependent. The control quantity is in this case the ratio $T_c/T_K$, which
is adjusted in experiments by changing the magnetic impurity species and
concentration (within the limits set by the miscibility of metals) or by
changing the doping $x$ in the Kondo superconductor systems such as
(La$_{1-x}$Ce$_{x}$)Al$_2$.

As previously described, we compute the thermodynamic quantity
$\braket{O}(T)$ at temperature $T$ from the NRG results for a fixed value of
gap $\Delta=\Delta(T)$. The results of such a calculation for a range of the
$T_K/T_c$ ratios are shown in Fig.~\ref{fig_c}. For $T \gtrsim T_c$, the
results are equivalent to those for a magnetic impurity in a normal metal.
For $T<T_c$, the gap starts to open. On a logarithmic scale the features in
thermodynamic quantities appear sharp (cusp-like), but a close-up reveals
that the change of slope is in fact continuous, albeit rapid. The results
are qualitatively equivalent to those obtained for fixed $\Delta$; a notable
difference, however, is the absence of the $\ln 3$ plateau. The plateau is
replaced by a rounded peak, since the condition of exact degeneracy of the
three many-particle states is satisfied only at a single temperature. The
peak in the entropy is then reflected in a characteristic temperature
variation of the impurity heat capacity, which has a positive and negative
part: this is the characteristic signature to be sought after in
experiments. We emphasize that the impurity heat capacity may be negative,
since it is defined as the difference of two total heat capacities (which
themselves are, of course, positive quantities).

\section{Discussion and conclusion}

We find that in Kondo superconductor systems an anomaly appears in the
impurity heat capacity for $T_K \sim T_c$. This is a large effect, since the
overall variation in the impurity heat capacity is of the order of $k_B$ and
it occurs in a temperature interval of approximately one decade of the
temperature scale. The positive part of the anomaly occurs at low
temperatures, i.e. in the exponential tail of the BCS heat capacity, thus
the heat-capacity enhancement is sizable and should easily be observable
already for impurity concentrations even below one permil. The negative part
of the anomaly occurs in the temperature range where the heat capacity of a
superconductor is enhanced, thus the heat-capacity reduction would be more
difficult to measure. Nevertheless, the magnitude of the effect ($0.1-0.2
k_B$ per impurity) is large enough that this effect could be detected at
somewhat higher impurity concentration in the permil range.

At high doping levels, the simple picture of independent impurities is no
longer valid due to interactions between the impurities. Owing to their pair
breaking properties, magnetic impurities lead to a suppression of the
superconducting transition temperature $T_c$ \cite{abrikosov1960,
ginsberg1974, zmhb, chung1996, balatsky2006}. In both the Abrikosov-Gorkov
treatment \cite{abrikosov1960} with classical impurity spins and the
Nagaoka-Suhl-approximation treatment by Zittarz and M\"uller-Hartmann
\cite{zmha, zmhb} with quantum impurity spins, the decrease of the critical
temperature is found to be proportional to the concentration divided by $k_B
\rho(0)$, where $\rho(0)$ is the density of states at the Fermi level in a
pure normal-state metal. For impurity concentrations in the permil range,
elemental superconductors with relatively high $T_c$ such as lead or niobium
are still superconducting. Another class of systems where these effects may
be detected are Kondo superconductor alloys. In fact, excess specific heat
at low temperatures had already been observed in
(La$_{1-x}$Ce$_{x}$)Al$_{2}$ alloy long ago \cite{armbruster1974}. Given the
progress in both experimental and theoretical techniques, it would
be interesting to revisit this problem focusing on the low-doping regime to
detect the predicted reduction of the heat capacity before the onset of the
enhancement and to look for possible semi-quantitative agreement. We
conclude by observing that a magnetic impurity in a bath may be considered
as an open quantum system thus this problem is also of interest in the
context of thermodynamics of nanosystems \cite{haenggi2008}.

\begin{acknowledgments}
We are much indebted to prof. K. Winzer for discussions on the subject of
Kondo superconductors. We acknowledge computer support by GWDG and support
by the German Science Foundation through SFB 602.
\end{acknowledgments}

\bibliography{paper}

\end{document}